\documentclass[aps,letterpaper,superscriptaddress]{revtex4}
\usepackage{amstext,amsmath,amssymb,ulem,amsthm}
\usepackage{color}

\usepackage{times}
\usepackage{algorithm}
\floatname{algorithm}{Protocol}
\usepackage{ftnxtra}
\usepackage{mwe}
\usepackage{subfig}

\usepackage[T1]{fontenc}
\usepackage[utf8]{inputenc}
\usepackage{graphicx}
\newcommand{\beq}{\begin{equation}}
\newcommand{\eeq}{\end{equation}}

\newcommand{\ket} [1] {\vert #1 \rangle}
\newcommand{\bra} [1] {\langle #1 \vert}
\newcommand{\braket}[2]{\langle #1 | #2 \rangle}

\newcommand{\tr}{\mathop{\mathrm{tr}}}

\newcommand{\ba}{\begin{align}}
\newcommand{\ea}{\end{align}}
\newcommand{\bea}{\begin{eqnarray}}
\newcommand{\eea}{\end{eqnarray}}

\newcommand{\R}{\mathbb{R}}

\newcommand{\norm}[1]{\left\lVert{#1}\right\rVert}

\newcommand\bsy[1]{\ensuremath{\boldsymbol{#1}}}

\usepackage{bm}
\usepackage{mathtools}

\usepackage{graphicx}  
\usepackage{dcolumn}   
\usepackage{bm}        
\usepackage{color}
\usepackage[]{algorithm}
\include{Qcircuit}
\usepackage{braket}
\usepackage{bbm}
\usepackage{mathtools}
\usepackage{amstext,amsmath,amssymb,ulem,amsthm}
\usepackage{color}

\usepackage{times}
\usepackage{algorithm}
\floatname{algorithm}{Algorithm}
\usepackage{ftnxtra}

\newcommand{\Ord}[1]{\mathcal{O}\left(#1\right)}
\newcommand{\Ordmax}[1]{\tilde{\mathcal{O}}(#1)}
\renewcommand{\braket}[2]{\left< #1 \left|#2 \right.\right>}

\newcommand{\matA}{\mathbf{A}}

\newcommand{\vecs}{\mathbf{s}}

\newtheorem{theorem}{Theorem}
\newtheorem{proposition}{Proposition}

\usepackage[T1]{fontenc}
\usepackage[utf8]{inputenc}
\usepackage{graphicx}
\usepackage[colorlinks]{hyperref}

\def\id{I}

\def\1{\mat{\id}}
\def\mat#1{\mathbf{#1}}

\renewcommand{\sout}[1]{}

\newcommand{\trc}[2]{\mathrm{tr}_{#1} \left\lbrace #2 \right\rbrace }

\begin{document} 

\title{Quantum-classical truncated Newton method for high-dimensional energy landscapes}

\author{Leonard Wossnig}
\email{leonard.wossnig.17@ucl.ac.uk}
\affiliation{Department of Computer Science, University College London, Gower Street, London WC1E 7JE, United Kingdom}
\affiliation{Institut f\"ur theoretische Physik, ETH H\"onggerberg, CH-8093 Z\"urich, Switzerland}
\affiliation{Department of Materials, University of Oxford, Parks Road, Oxford OX1 3PH, United Kingdom}

\author{Sebastian Tschiatschek}
\affiliation{Department of Computer Science, ETH H\"onggerberg, CH-8093 Z\"urich, Switzerland}

\author{Stefan Zohren}
\affiliation{Department of Materials, University of Oxford, Parks Road, Oxford OX1 3PH, United Kingdom}

\date{\today}

\begin{abstract}
We develop a quantum-classical hybrid algorithm for function optimization that is particularly useful in the training of neural networks since it makes use of particular aspects of high-dimensional energy landscapes. Due to a recent formulation of semi-supervised learning as an optimization problem, the algorithm can further be used to find the optimal model parameters for deep generative models. In particular, we present a truncated saddle-free Newton's method based on recent insight from optimization, analysis of deep neural networks and random matrix theory. By combining these with the specific quantum subroutines we are able to exhaust quantum computing in order to arrive at a new quantum-classical hybrid algorithm design. Our algorithm is expected to perform at least as well as existing classical algorithms while achieving a polynomial speedup. The speedup is limited by the required classical read-out. Omitting this requirement can in theory lead to an exponential speedup.
\end{abstract}

\maketitle

 
\section{Introduction}

In the literature of quantum machine learning many fundamental machine learning algorithms can be found~\cite{ciliberto2017quantum}, including quantum support vector machines \cite{rebentrost2014quantum}, k-means clustering \cite{lloyd2014quantum,wiebe2015quantum}, variants of linear regression \cite{wiebe2012quantum,liu2017fast,prakash2014quantum,schuld2016prediction}, Bayesian inference \cite{low2014quantum,wiebe2015can}, Gaussian processes \cite{zhao2015quantum}, perceptron models \cite{wiebe2016quantum}, and principal component analysis \cite{lloyd2014quantum}. However, particularly in big data scenarios, data sets often contain only few labeled samples, while most acquired samples are unlabeled. Many traditional methods do not generalize well in this case, resulting in bad predictions on unseen data. The reason for labelling only small fractions of the available data is usually the high cost -- labelling data often requires lots of expertise and time. Semi-supervised learning addresses this problem by using large amounts of unlabelled data together with the labeled data, to build classifiers which are able to generalize to unseen data. Hence, it is important to dedicate efforts to designing good models, features, kernels or similarity functions for semi-supervised learning~\cite{zhu2006semi}. Using the framework of Kingma and Welling~\cite{kingma2013auto}, any quantum algorithm for optimization in high-dimensional energy landscapes can be applied to train deep auto-encoders for semi-supervised learning. This is one of the motivations for introducing quantum-classical algorithms for optimization in high-dimensional energy landscapes.

The idea of quantum machine learning is to use the inherent properties of quantum mechanics to improve or speedup the process of algorithmic learning.
The majority of the existing algorithms are based on the replacement of classical subroutines for linear algebra or search (which is similar to sampling) with a quantum equivalent albeit an expected-to-be faster algorithm.

The most common approach in quantum computing is to represent classical data/information as binary strings which can be directly translated into $n$-qubit quantum states $\ket{x_1 x_2 \ldots x_n}$
with basis $$\left\lbrace \ket{00 \ldots 0},\ket{00\ldots1},\ldots,\ket{11\ldots1} \right\rbrace$$ in a $2^n$ dimensional Hilbert space or storing the values required for computation in the amplitude of a state, e.g.\ encoding a matrix in a quantum state that takes the form $\ket{A}= \sum_i \sum_j \alpha_i^j \ket{i} \ket{j}$.\
The information can be read out through measurements of the single bits which requires time at least $\Ord{N}$ due to the stochastic nature of the measurement process and the
fact that per measured qubit only one bit of information can be retrieved. Hence one important subtlety of quantum machine learning is the requirement of classical outputs, where in order to maintain efficiency we need to strongly reduce
the possible outcomes of our simulation prior to the measurement. Finding genuinely quantum ways to represent and extract the information is here the key to successful quantum speedup.

In this work we present a classical-quantum hybrid algorithm for optimizing a function in a high-dimensional space. Classical first order methods run in linear time and are considered to be efficient. However, for very high-dimensional, non-convex energy landscapes such methods are known to perform poorly~\cite{martens2010deep}. We thus focus on second order methods. In particular, we develop a quantum algorithm for truncated saddle-free Newton's method, an adapted iterative second order method.
We first motivate the need for these second order, curvature sensitive methods in section~\ref{sec:second_order_methods} and then, after a short review of existing methods in section~\ref{sec:comparison}, present the main result of this work in section~\ref{sec:quantum_truncated} which is a hybrid algorithm for an adapted Newton's method which builds upon recent results in quantum information processing~\cite{harrow2009quantum,lloyd2014quantum,rebentrost2014quantum,schuld2016prediction}, a number of empirical insights from training of deep neural networks~\cite{martens2010deep,pascanu2014saddle,dauphin2014identifying,sagun2016singularity,sagun2017empirical} as well as random matrix theory~\cite{laloux2000random,bloemendal2016principal}.

We describe a general method for performing a step of a truncated version of the saddle-free Newton's method~\cite{dauphin2014identifying,arjovsky2015saddle} and reinforce our theoretical argument with empirical evidence for different data sets. In order to perform a step, density matrices that encode the Hessian matrix are simulated using a known sample based Hamiltonian simulation scheme~\cite{lloyd2014quantum} and oracles which allow the efficient access to these are assumed. We then use specific problem traits which have been empirically observed in deep Neural Networks and perform a classical truncated Newton's step. This work represents hence a new approach to designing quantum-classical hybrid algorithms based on a combination of empirical and theoretical insights from classical optimization theory to design new algorithms which can harness specific problem traits by using features of quantum computation. Our algorithm scales better than existing classical algorithms but only offers a polynomial speedup as we require a classical output. If this restriction is omitted the algorithm may obtain a exponential speedup. We pose it as open question whether our method can be extended to a purely quantum algorithm in the spirit of the work by Rebentrost et al.~\cite{rebentrost2016quantum} or Kerenidis and Prakash~\cite{kerenidis2016quantum}.

In addition to providing a novel quantum-classical hybrid algorithm, some of the ideas behind the construction of the algorithm also provide some new perspective on second order methods for optimisation in the context of neural networks and relations to random matrix theory.

\section{Preliminaries}

We first present a few linear algebra preliminaries and then review preliminaries of quantum information theory.

\subsection{Spectral analysis of matrices} 

The set $\{1,2,\ldots,n\}$ is denoted by $[n]$, and the standard basis vectors in $\R^n$ are denoted by $e_i, \, i \in [n]$.
For any matrix $A \in \R^{m \times n}$ the spectral norm is defined as $||A||_2 = \sqrt{ \lambda_{max}(A^{\dagger}A)} = \sigma_{max}$ and the Frobenius norm of $A$ is defined as $||A||_F = (\sum_{ij} |A_{ij}|^2)^{1/2} = (\sum_{i}^{\text{min}(m,n)} \sigma_i)^{1/2}$, where $\sigma_i$ are the singular values of $A$. For a vector $\mathbf x$ we denote the norm $||\mathbf x ||^2 = \sum_i  x_i^2$. A matrix $A$ is unitary if $AA^{\dagger} = A^{\dagger}A = I$. A matrix $P \in \R^{n \times n}$ is a projector if $P^2=P$ and has eigenvalues $0$ or $1$. If $A$ is a matrix with orthonormal columns, then $AA^{\dagger}$ is a projector into column space of $A$.\\

We further need the following well known results from linear algebra.

\begin{theorem}[Singular Value Decomposition (SVD)]
For $A \in \mathbb R ^{m \times n}$ there exists a decomposition of the form 
$A = U \Sigma V^{\dagger}$, where $U \in \mathbb{R}^{m \times m}$ ,$V \in \mathbb{R}^{n \times n}$ are unitary operators, and $\Sigma\in \mathbb{R}^{m \times n}$ is a diagonal matrix 
with $r$ positive entries $\sigma_1 ,\sigma_2, \ldots, \sigma_r$, and $r$ is the rank of $A$. As alternative notation, we write $A=\sum_{i}^{r}\sigma_i\mathbf{u}_i\mathbf{v}_i^{\dagger}$, where $\{\mathbf{u}_i\}$, $\{\mathbf{v}_i\}$, and $\{\sigma_i\}$ are the sets of left and right mutually orthonormal singular vectors, and singular values of $A$ respectively.
\end{theorem} 
The Moore-Penrose pseudo-inverse of a matrix $A$ with singular value decomposition as defined above is then given by $A^+ = V \Sigma ^{+} U^{\dagger} = \sum_{i}^{r}(1/\sigma_i)\mathbf{v}_i \mathbf{u}_i^{\dagger}$. The matrix $AA^+$ is the projection onto the column space $Col(A)$ while $A^+A$ is the projection onto the row space $Row(A)$.

\begin{theorem} [Spectral Decomposition (SD)]
For $A \in \mathbb R ^{n \times n}$ there exists a decomposition of the form 
$A = S \Lambda S^{\dagger}$ where $S \in \mathbb{R}^{n \times n}$ is a unitary operator, and $\Lambda=\text{diag}(\lambda_1, \ldots, \lambda_n)$ is a diagonal matrix whose entries are the eigenvalues of $A$. We write $A=\sum_{i}^{n}\lambda_i\mathbf{s}_i\mathbf{s}_i^{\dagger}$, where $\{\mathbf{s}_i\}$ and $\{\lambda_i\}$ are the sets of eigenvectors and eigenvalues of $A$ respectively.
\end{theorem} 
A useful relationship between the singular values and eigenvalues is given by the following proposition.
\begin{proposition}
  \label{prop1}
  If $A=A^{\dagger}$ is a hermitian matrix, then the eigenvalues are equal to the singular values of $A$ up to a sign ambiguity and the corresponding
  eigenvectors are equal to the singular vectors of $A$, i.e.\ $\sigma_i = |\lambda_i|$ and $\mathbf{u}=\mathbf{v}=\mathbf{s}$. Therefore the singular value decomposition and the spectral decomposition are related by $A=\sum_{i}\lambda_i\mathbf{s}_i\mathbf{s}_i^{\dagger} = \sum_i \pm \sigma_i\mathbf{u}_i\mathbf{u}_i^{\dagger}$.
\end{proposition}
We use this connection to obtain the eigenvalues of a hermitian matrix $A$ and apply an approximation of the inverse Hessian to the gradient.

\subsection{Random matrix theory} 

Some of the steps of the algorithm presented below are motivated by random matrix theory and we review the some preliminaries here. Let $A \in \mathbb R ^{m \times n}$ be a $(m \times n)$-dimensional matrix with all elements independently and identically distributed according to a normal distribution $\mathcal{N}(0,\sigma^2)$ of mean 0 and finite variance $\sigma^2$. Then the matrix $W= n^{-1}A A^T$ is a $(m\times m)$-dimensional random matrix known as a Wishard matrix. In the limit of infinite size the density of eigenvalues of a Wishard matrix, formally defined as $\mu(J) = \lim_{m\to\infty} \mathbb{E} (\text{number of eigenvalues in $J$})/m$ for $J\subseteq \mathbb{R}$, is given by the so-called Marchenko–Pastur distribution (see Figure \ref{fig:MP}):

\begin{figure}
\begin{center}
  \includegraphics[width=5in]{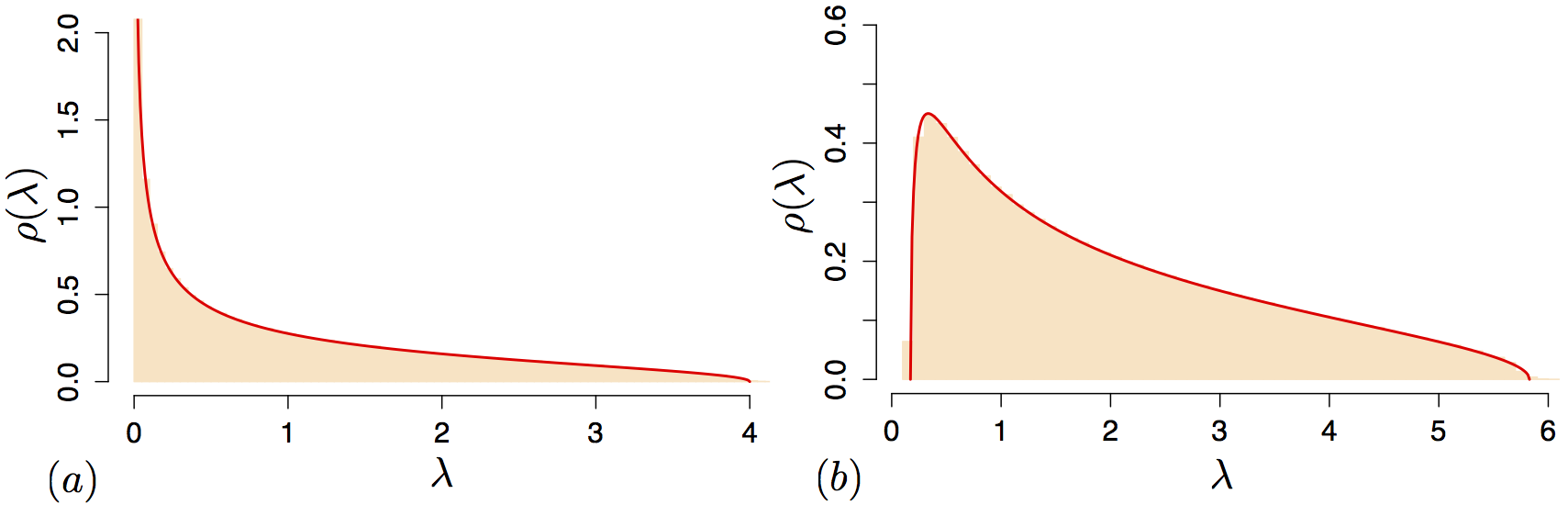}
  \caption{Illustration of the Marchenko–Pastur distribution for (a) $c=1$ and (b) $c=1/2$ with $\sigma^2=1$. In addition to the analytical expression (red) we also show the histogram of eigenvalues obtained from 1000 samples of $W$ with sizes $n=100$ and (a) $m=100$ or (b) $m=50$.}
  \label{fig:MP}
  \end{center}
\end{figure}

\begin{theorem} [Marchenko–Pastur distribution \cite{marvcenko1967distribution}] For $m,n\to\infty$ and $m/n\to c > 0$ one has
\begin{eqnarray}
\mu(J) = \begin{cases}
\nu(J)     &  0<c \leq 1, \\
 \nu(J) + \left(1-1/c\right) 1_{J\ni 0}   & 1< c.
\end{cases}
\end{eqnarray}
with 
\begin{eqnarray}
d\nu(\lambda) =: \rho(\lambda) d\lambda = 
 \frac{1}{2\pi\sigma^2} \frac{\sqrt{(c_+-\lambda)(\lambda - c_-)}}{c\lambda} 1_{\lambda \in (c_-,c_+)} d\lambda,
\end{eqnarray}
with $c_\pm = \sigma^2(1\pm\sqrt{c})^2$.
\end{theorem} 

The theory of random matrices and the Marchenko–Pastur distribution in particular are important when analysing the Hessian of the loss function of deep neural networks as we discuss in more detail in the next section. While the above results only hold true for random (Wishard) matrices, we will see that when adding some correlation structure part of the spectrum still follows the Marchenko–Pastur distribution.

\subsection{Phase estimation and state preparation}
\label{sec:pe_and_sp}

Our algorithm relies on the well known quantum phase estimation procedure which was introduced by Kitaev~\cite{kitaev1995quantum} and is summarized in the following theorem:
\begin{theorem}[Phase estimation \cite{kitaev1995quantum}]
  \label{pest}
  Let $U$ be a unitary, $U \ket{v_j} = \exp{(i \theta_j)} \ket{v_j}$ with $\theta_j \in [ - \pi ,\pi ]$ for $j \in [n]$. There is a quantum algorithm that transforms $\sum_{j \in [n]} \alpha_j \ket{v_j} \to \sum_{j \in [n]} \alpha_j \ket{v_j} \ket{\overline{\theta}_j}$ such that $|\overline{\theta_{j}} - \theta_{j} |\leq \delta$ for all $j\in [n]$ with probability $1-1/\text{poly}(n)$ in time $\Ord{T_U \log{(n)} / \delta}$, where $T_U$ defines the time to implement $U$.
\end{theorem}

In order to create the states in our algorithm we need to be able to prepare the density matrix that encodes the square of the Hessian.
We can then use the eigenvalues of this matrix to apply the inverted Hessian to the gradient which is well known in the quantum algorithm literature~\cite{harrow2009quantum,schuld2016prediction}.
For our algorithm to work we assume the following oracles:
\begin{eqnarray}
 U_f \ket{\mathbf x} \ket{i,j} \ket{0}= \ket{i,j}\ket{H_{ij}} \\
 U_g \ket{\mathbf x} \ket{i} \ket{0}= \ket{i}\ket{(\nabla f)_i} ,
\end{eqnarray}
where the oracle can -- when called with input values $i,j$ or $i$ respectively --  prepare a state that contains the $i,j$-th values of the Hessian matrix or $i$-th value of the gradient evaluated at point $\mathbf x$.
This assumption is made based on the fact that the gradients and Hessian matrix can in many cases be calculated via an explicit formula, so that we can construct an oracle for it. Note that solely the task of calculating and storing the Hessian matrix for high-dimensional spaces is infeasible in the classical case for high-dimensional inputs. For example this issue arises in the training of deep neural networks. Here the quantum method could be a suitable application.
For the case that the function can only be evaluated numerically the gradient can simply be calculated in superposition
using finite differences and an oracle to evaluate the function in superposition.
Since the Hessian is symmetric, it holds that $H_{ij}=H_{ji}$. We use this in the preparation of the density matrix by preparing a superposition of all states $i,j$ and query the oracle $U_{g/f}$ to obtain the entries of the Hessian and gradient in superposition in time $T(U_{g/f})$. 
In Appendix~\ref{app:density_matrix} we will outline the procedure to obtain the density matrix in detail.
By following this procedure we are able to prepare copies of the state, c.f.\ \eqref{final_density},
\begin{eqnarray}
    \rho_{HH^{\dagger}} &=&\sum_{i,i',j}\alpha _{i,j}\alpha
    ^*_{i',j}\frac{H_{ij}H_{ji'}}{C^2}|i\rangle \langle i'|,
\end{eqnarray}
where $C$ is a proper normalization constant.
We need multiple copies of this density matrix in the matrix-exponentiation step of our algorithm.

\section{Motivation and previous work}

\subsection{The need of second order methods}
\label{sec:second_order_methods}

Second order methods are not the first choice in practical optimization, since they usually bear too large computational costs. However, for certain problems they offer fundamental advantages compared to first order methods, and allow us to optimize objective functions which are restrictive for first order methods. One classical example is the Rosenbrock function. First order methods like gradient descent (GD) algorithms perform well for shallow networks and are efficient since the gradients can be calculated efficiently. Unfortunately for deep neural networks it is well known that these methods do not generalize very well, a problem that is commonly known as the \textit{vanishing gradient problem}. GD is in particular unsuitable for optimizing objectives with a pathological curvature, which is a well known fact in the optimization community~\cite{martens2010deep}.
Here second order methods which correct the steps according to the local curvature have been demonstrated to be more successful. Martens~\cite{martens2010deep} argues that deep networks exhibit such pathological curvature, which makes curvature-blind methods likely to fail. Second order methods in contrast are infeasible for large models due to the quadratic relationship between the number of parameters (i.e. weights and biases in a MLP) and the Hessian matrix. 
This is the reason why many algorithms have been developed that try to approximate these. They all try to use second order information to be able to approximate the Newton step. Newton's method is an extension of the standard gradient descent algorithm which iteratively performs updates of the current solution to converge to a local optimum. The update rule of gradient descent for an objective function $f:\mathbb R^N \rightarrow \mathbb R$ and $x \in \R^N$ is given by
 \begin{eqnarray}
 \label{eq:grad_desc}
 x^{t+1} = x^t - \eta \nabla_x f(x) |_{x=x^t},
 \end{eqnarray}
 where the parameter $\eta$ is the step-size or learning rate, which determines the convergence of the algorithm and usually one uses a Heuristic to adaptively set it.
The behaviour of these methods can be analysed using the formulation of Morse’s lemma~\cite{dauphin2014identifying} for non-degenerate saddle points for which the Hessian is not singular. Let $f$ be the function we want to optimize. The behavior around the saddle-point can then be seen by re-parameterizing the function according to:
\begin{eqnarray}
	f(\mathbf x^* + \Delta \mathbf x) = f(\mathbf x^*) + \frac{1}{2} \sum\limits_i \lambda_i \mathbf v_i^2,
\end{eqnarray}
where the $\lambda_i$ represent the eigenvalues of the Hessian and the $\mathbf v_i$ are the new parameters of the model corresponding to the motion along the eigenvector $\mathbf e_i$ of the Hessian of $f$ at point $\mathbf x^*$.
Newton's method adopts the stepsize in each direction of the Hessian by rescaling the gradient descent step, which is given by $-\lambda_i \mathbf v_i$. This yields a step $-\mathbf v_i$. However, for a negative eigenvalue this results in a move in opposite direction to the gradient descent direction. This means that a saddle point, where some directions have a positive and some have a negative eigenvalue, becomes an attractor for Newton's method. If, instead of dividing by the eigenvalue $\lambda_i$, one takes the absolute value $|\lambda_i|$ when rescaling the gradient step, one obtains the saddle-free Newton's method~\cite{dauphin2014identifying}. This allows to take a step in the correct direction while still maintaining the corrected step-size. Although this theoretical framework is a good visualization of the concept of Newton's method, it doesn't justify the usage of saddle-free Newton's method (i.e.\ renormalization with the absolute of the eigenvalues) in general. In particular as the Hessian is singular at most saddle points~\cite{sagun2016singularity}. However, Dauphin~\cite{dauphin2014identifying} demonstrated that their approach is indeed valid in a more general setting. Thus we find that saddle-free Newton's method allows to optimize a wide range of objectives and if implemented efficiently constitutes a highly valuable tool for the training of deep neural networks.

\subsection{Review of existing approaches}
\label{sec:comparison}

Dauphin et al.~\cite{dauphin2014identifying} introduced a low-rank approximation scheme for the saddle-free Newton's method which takes time $\Ord{k N^2}$, using a rank-$k$ approximations to the Hessian matrix based on fast evaluation of the $k$ Laczos vectors, and hence performing the saddle-free optimization in the Krylov-subspace. We note that there exists a classical sampling-based (randomized) algorithm which allows us to obtain a low-rank approximation in $\Ord{k^2N}$ time, which can be used in this setting as well. However, this method has very weak bounds on the error of the resulting low rank approximation, which scales with the norm of the matrix we want to approximate and inverse linearly with the sample-size. This result is briefly reviewed in the appendix~(\ref{app:sample_svd}). In comparison~\cite{arjovsky2015saddle} proposes a saddle-free Hessian-free algorithm that takes time $\Ord{m \cdot l \cdot N}$ time where $m$ is the time required for the solution of an ordinary differential equation and $l$ is the number of steps required for the conjugated gradient. The latter argues that the number of required principal components $k$ in the method given by~\cite{dauphin2014identifying} becomes prohibitively large in order to get a useful approximation, i.e.\ one which captures a high percentage of the total variance of the data, especially for the memory cost which is of $\Ord{kN}$. Furthermore~\cite{arjovsky2015saddle} argues that $m$ is typically small for random matrices and hence conjectures that this number will be small in the algorithm. Despite these methods, a variety of approximation methods, like quasi-Newton's method~\cite{haelterman2009analytical} exist, which give good results as well. However, there is no general consensus on one particular method working \textit{best in practice}, and performance depends strongly on the particular problem. Hessian-free methods~\cite{martens2010deep} use the Newton direction (which is asymptotically quadratically convergent), but require to compute several Hessian-projections, each at approximately the cost of a gradient evaluation. Quasi-Newton~\cite{haelterman2009analytical} methods in comparison only approximate that direction weakly and thereby obtain a super-linear instead of the quadratic scaling.
When comparing the different second-order approximative methods, it is not clear whether investing into a locally good direction (Hessian-free), or moving aggressively in a sub-optimal direction (quasi-Newton) is more beneficial. This depends furthermore on the individual objective in practical applications.

\section{Quantum algorithm for truncated Newton's method} \label{sec:quantum_truncated}

We describe general methods for finding the minimum of such an objective function involving deep neural networks. Depending on the form of the objective function one can either obtain the solution by directly solving a linear system (for least squares objective) or by using iterative methods like gradient descent. 

Here we present an iterative quantum-classical hybrid solver to obtain the minimum of such an objective function which uses insights from practical considerations from classical optimization tasks as well as the eigenvalue distributions of large neural networks and random matrix theory. This allows us to construct an iterative algorithm which is widely applicable, expected to converge faster and has a better than classical dependence on the dimensionality of the input, i.e.\ the number of parameters in the neural network. In order to perform one step of this algorithm we use the method of sample based Hamiltonian simulation~\cite{lloyd2014pca} to exponentiate density matrices that encode the Hessian matrix. We assume oracles that allow the efficient access to the gradient and Hessian which can be implemented for example based on Jordan's quantum algorithm for numerical gradient estimation~\cite{jordan2005fast} or a simple calculation of the different gradients in superposition followed by a controlled rotation and post-selection. The latter appears to be a realistic assumption, given that one can derive an analytic description of the gradient and Hessian matrix for most of the models which are used. Based on these components, the matrix inversion for the Hessian matrix can be performed via well-known techniques (e.g.~\cite{harrow2009quantum}). In the following we first argue for the neccessity of second order methods, then present our algorithm and finally discuss the validity of this approach. 
The argumentation is based on results of numerical experiments. Additionally to our own results, we can refer to older results of the classical saddle-free Newton method and similar approaches~\cite{dauphin2014identifying,arjovsky2015saddle}.

Let $N$ denote the dimension of the parameter vector, i.e. $N$ is related to the dimension of the matrix that represents the neural network e.g.\ $N = l^k $ for $k$ layers with $l$ neurons with all-to-all connections.
The algorithm we present can perform one step of the classical saddle-free Newtons method~\cite{dauphin2014identifying}, i.e. approximately apply the absolute inverse of the Hessian matrix, where the absolute of the matrix $\matA$ with eigenvalues $\lambda_i$ and eigenvectors $\vecs_i$ is defined as $\norm{\matA} = \sum_i \lambda_i \vecs_s \vecs_i^T$. to the gradient and read-out the update step in $\Ord{N \log{N}}$ time. The main contribution to this runtime is a result of the requirement to write down the quantum state in each step, which takes $\Ord{N \log N}$, since we need to repeatedly measure the quantum state to obtain the full vector for the update direction. We leave it as an open question whether this can be extended to the pure quantum algorithm, similar to the one given by Rebentrost et al.~\cite{rebentrost2016quantum}.
 Let $k$ denotes the number of singular values used in the approximation of the algorithm which is discussed in detail later. Depending on how many steps the algorithm are required to converge, we obtain a final scaling of $\Ordmax{\mathrm{iter} \cdot k \cdot N \log{N}}$), where we expect $\mathrm{iter}$ to be substantially smaller than in the classical stochastic gradient descent (SGD) for deep neural networks. In particular this runtime is  better than the classical runtime of the saddle-free Newtons method~\cite{dauphin2014identifying} with $k N^2$ and can obtain a speedup even compared to the $\Ord{k^2 N}$ runtime of the same algorithm which uses a probabilistic singular value decomposition. The Saddle-free Hessian-free method~\cite{arjovsky2015saddle} in comparison takes $\Ord{l m N}$, where $l$ is the number of steps taken by the conjugate gradient method and $m$ is the number of steps required by the ODE solver. Even here our algorithm potentially allows for a speed-up. Note that both algortihms, saddle-free and Hessian-free saddle-free, use a cruder approximation since they perform an incomplete optimisation step.

The space requirements of our algorithm are of $\Ord{\log{N}}$, if we can evaluate the Hessian matrix through an explicit formula, which can be beneficial compared to classical requirements.
As we discussed above, we assume existence of an oracle which allows the access to the data which has a similar form as the Hessian and Gradient oracles described above. This is a standard assumption in many quantum machine learning algorithms (see for example ~\cite{lloyd2014quantum}). We can now investigate the runtime of our algorithm in the query setting. Let $T(U)$ be the time to access the oracles. Then our algorithm can take one step, using this oracle in time $\Ord{T(U)\cdot N \cdot \text{poly}(\log{N})}$, which reduced to $\Ord{N \cdot \text{poly}(\log{N})}$ if the given oracles are efficiently implementable, i.e. that $T(U)$ can for example be $\text{poly}(\log{d \cdot r})$ assuming a form of quantum memory architecture which allows access to the $r$ data points with dimension $d$ in quantum parallel. Without quantum parallel access to the data or the Hessian matrix respectively, the potential speedup of our algorithm is lost. This however is a known restriction for most known quantum (machine learning) algorithms.

  In the following we describe the algorithm.
Assume that the first and second order derivatives can be calculated and there exists a function that
when called with the indices $i$ or $(i,j)$ respectively, and some input register $\ket{\mathbf x}$, 
can return the values of the gradient and Hessian evaluated at $\mathbf x$.

\begin{enumerate}
  \item We require many copies of the state described in eq.~(\ref{final_density}) in the Appendix in order to obtain a density matrix that encodes the Hessian matrix times its transpose. We can obtain many copies of the state
    \begin{eqnarray}
      \rho _{HH^{\dagger}} &=&\sum_{i,i',j}\alpha _{i,j}\alpha
      ^*_{i',j}\frac{H_{ij}H_{ji'}}{c^2}|i\rangle \langle i'|
    \end{eqnarray}
    using the mentioned oracles. We then perform a Hamiltonian Simulation with these matrices, based on the well-known method of density matrix exponentiation~\cite{lloyd2014pca}. 
    We note that for our purpose this method is sufficient since we will deal with low-rank matrices as we will show below. In a more general case one could imagine to use standard methods for Hamiltonian simulation ~\cite{harrow2009quantum} or similar if the Hessian is stored in memory then methods similar to~\cite{wossnig2017quantum} can be applied.

  \item Let $\tilde{\sigma}$ be the time evolution we obtain by the protocol. Then $||e^{-i \rho t } \sigma e^{i \rho t} - \tilde{\sigma} || \leq \epsilon$ is the error. Using $\Ord{t^2 \epsilon^{-1}} \equiv n$ copies of $\rho$ allows to implement $e^{-i \rho t}$ with error $\epsilon$ in time $\Ord{n \text{log}(N)}=\Ord{t^2 \epsilon^{-1} \text{log}(N)}$ with the technique presented in the quantum principal component analysis by Lloyd et al.~\cite{lloyd2014quantum}. This scaling for the exponentiation of the density matrix is essentially optimal~\cite{kimmel2017hamiltonian}. We then obtain the state:
  \begin{eqnarray}
    e^{-i \rho_{HH^{\dagger}} t} \ket{\chi}
  \end{eqnarray}
  where $\chi$ is the state that encodes the gradient, i.e.\ $\ket{\chi} = \ket{\nabla f}$.
  This is done by repeated application of the Swap-operator followed by tracing out the subsystem in the following manner:
  \begin{eqnarray}
            && {\tr}_1 \left( e^{-iS \Delta t} \rho \otimes \sigma e^{iS \Delta t} \right) \nonumber 
            = {\tr}_1 \left([ 1-iS \Delta t + \Ord{\Delta t^2} ] \rho \otimes \sigma [ 1+iS \Delta t + \Ord{\Delta t^2} ]  \right) \nonumber \\ 
            &=& {\tr}_1 \left( \rho \otimes \sigma  -iS  \rho \otimes \sigma \Delta t + iS \rho \otimes \sigma \Delta t + \Ord{\Delta t^2}  \right) \nonumber  
            = \sigma - i \Delta t \left[\rho, \sigma \right] + \Ord{\Delta t^2}
  \end{eqnarray}
  and note that
  \begin{eqnarray}
  \label{eq:density_exp_calculus}
    \sigma - i \Delta t \left[\rho, \sigma \right] + \Ord{\Delta t^2} \approx e^{-i \rho \Delta t} \sigma e^{i \rho \Delta t}
  \end{eqnarray}
  with $\sigma \equiv \ket{\chi}\bra{\chi}$. This can be seen by comparing the above equation~(\ref{eq:density_exp_calculus}) to
  \begin{eqnarray*}
    && e^{-i \rho \Delta t} \sigma e^{i \rho \Delta t} 
    = [ 1-i \rho \Delta t + \Ord{\Delta t^2} ] \sigma  [ 1+i \rho \Delta t + \Ord{\Delta t^2} ]
    = \sigma - i \Delta t \left[\rho, \sigma \right] + \Ord{\Delta t^2}.
  \end{eqnarray*}

  Since $H$ is a symmetric matrix, the eigenvectors and the singular vectors are equivalent and the singular values are up to a sign ambiguity
  equivalent to the eigenvalues, we can use that $HH^{\dagger}$ has eigenvalues $\lambda_i^2$ and eigenvectors $\ket{u_i}$ which coincide with the squared eigenvalues and eigenvectors of $H$. The decomposition of $\ket{\chi}$ in the eigenvector-basis of $H$ is denoted as $\sum_i \eta_i \ket{u_i}$.
  We then obtain:
   \begin{eqnarray}
    e^{-i \rho_{HH^{\dagger}}  t} \ket{\chi} = \sum_i \eta_i  e^{-i \lambda_i^2  t} \ket{u_i}.
  \end{eqnarray}

  \item We can recover the eigenvalues using the phase estimation algorithm with controlled-swap operations and we obtain the state
   \begin{eqnarray}
    \sum_i \eta_i  e^{-i \lambda_i^2  t} \ket{u_i} \ket{0} \overset{PE}{\longrightarrow} \sum_i \eta_i \ket{u_i} \ket{\overline{\lambda}_i^2}
  \end{eqnarray}
  where $\overline{\lambda}_i$ is the estimate of the true eigenvalue $\lambda_i$ which we obtain through phase estimation up to a certain accuracy.
  For a positive semidefinite $H$ we only need to continue with the common matrix inversion techniques as introduced by Harrow et al.~\cite{harrow2009quantum}. In effect we will thereby apply the pseudo-inverse of the Hessian. For a general matrix we would instead require to recover the sign of the eigenvalue at this stage, similar to the algorithm of Wossnig et al.~\cite{wossnig2017quantum}. 
  However, for our purpose we instead invert the singular values which are the absolute of the eigenvalues of the Hessian matrix directly. This implements the saddle-free Newton's method, also called the damped Newton's method, as introduced by Dauphin et al.~\cite{dauphin2014identifying}. This adjustment is necessary to allow the method to escape saddle points which are prevalent in high-dimensional spaces, in particular in the case of neural network training~\cite{dauphin2014identifying} as discussed above. 

  \item We then continue by applying the conditional rotation $R_y(2 \text{sin}^{-1}(1/|\lambda_i|))$, where $|\lambda_i|=\sqrt{\lambda_i^2}$.
  The rotation prepares the state
  \begin{eqnarray}
    \sum_i \eta_i \ket{u_i}_A \ket{\overline{\lambda}_i^2}_B \left( \sqrt{1-\frac{c}{|\overline{\lambda}_i|^2}} \ket{0}_C + \frac{c}{|\overline{\lambda}_i|} \ket{1}_C \right),
  \end{eqnarray}
  where $c=\Ord{1/\kappa}$ for the condition number $\kappa(H)$, and postselect on the rightmost qubit being in state $\ket{1}$. The probability of success of this step is at least $1/\kappa^2$, since $$p_{success} = \sum_i |\eta_i|^2 \cdot |C/\lambda_i|^2 = \sum_i |\eta_i|^2 \cdot |1/\kappa^2\lambda_i|^2 \leq 1/\kappa^2.$$ Using amplitude amplification this can be boosted to $1/\kappa$. We hence need $\Ord{\kappa}$ repetitions of this step on average for success. We thereby applied the pseudo-inverse of the absolute Hessian matrix to the gradient, i.e. $|H|^{-1}\nabla f$. 
  Since the eigenvalues are upper bound by $1$ \footnote{Since the eigenvalues of a density matrix sum up to $1$} but do not have a lower bound and can in fact be of $\leq \Ord{1/N}$ the condition number can be as large as $\Ord{N}$, which would result in an exponential runtime. We deal with this issue by using some insights from random matrix theory and then truncating the inversion with a certain threshold value using filter functions for the eigenvalues similar to the idea of filtering the ill-conditioned part of the matrix introduced by Harrow et al.~\cite{harrow2009quantum}.

  First, we will review some details about low-rank approximations of a matrix $H$.
  We start with the matrix $H= \sum_{i=1}^N \sigma_i u_i v_i^{\dagger}$. The best $k$-rank approximation
  is then given by the Eckart–Young–Mirsky theorem ~\cite{eckart1936approximation} 
  $$H_k =  \sum_{i=1}^k \sigma_i u_i v_i^{\dagger}.$$
  Since we deal here with positive semi-definite matrices, this tranlates to
  \begin{eqnarray}
    H =  \sum_{i=1}^N \sigma_i u_i v_i^{\dagger} = \sum_{i=1}^N \lambda_i s_i s_i^{\dagger} , \\
    H_k =  \sum_{i=1}^k \sigma_i u_i v_i^{\dagger} = \sum_{i=1}^k \lambda_i s_i s_i^{\dagger},
  \end{eqnarray}
  where we assume the ordering of the singular and eigenvalues from the largest to the smallest $\sigma_1 \geq \sigma_2 \geq \ldots$.
  Therefore we can see that only taking into account the $k$ largest singular values is similar to a $k$-rank approximation. This translates then 
  into an error of
  \begin{eqnarray}
    \Delta H_k = || H - H_k ||_2 = || \sum_{i=k+1}^N \lambda_i s_i s_i^{\dagger} ||_2 = \lambda_{k+1}
  \end{eqnarray}
  or if we consider the Frobenius norm instead
  \begin{eqnarray}
    || H - H_k ||_F = || \sum_{i=k+1}^N \lambda_i s_i s_i^{\dagger} ||_F = \sum_{i=k+1}^N \lambda_{i}.
  \end{eqnarray}
  which is the best $k$-rank approximation for the matrix $H$. 
 
  For a few large eigenvalues and a bulk of small ones with several orders of magnitude difference the approximation will give a good representation if we discard the eigenvalues at a proper cutoff. The resulting error 
  will then as low as $\Ord{1/N}$. If all eigenvalues are of a similar order the relative error will still be large, since that $\Delta H_j/ \Delta H_{jk} = \Ord{1}$ for all $j,k$,
  since all eigenvalues are then of $\Ord{1/N}$.
  
  Since we want to apply the inverse of the Hessian, given by 
  \begin{equation}
    H^{-1} = \sum\limits_{i=1}^N \lambda_i^{-1} s_i s_i^{\dagger} = \sum\limits_{i=1}^N \tilde{\lambda}_i s_i s_i^{\dagger},
  \end{equation}
  at first glance one might believe that we need to approach the problem differently. However, Chung et al. [Theorem 3 in~\cite{chung2015optimal}] showed that the optimal low rank approximation of the inverse, which is the unique minimizer, is also given by the largest $r$ singular values, i.e. the optimal low rank approximation $Z^*$ of $H^{-1} \in \mathbb{R}^{N\times N}$
  \begin{eqnarray}
    Z ^*= \underset{Z:\text{rank}(Z) \leq r}{\text{argmin}} || Z H - Id_N||_F^2,
  \end{eqnarray}
  for $H = U \Sigma V^{\dagger}$ is given by 
  \begin{eqnarray} \label{inverserank}
    Z^* = V_r \Sigma_r U_r^{\dagger},
  \end{eqnarray}
  where $\Sigma_r = \text{diag}\left(\frac{1}{\sigma_1},\ldots,\frac{1}{\sigma_r} \right)$ and $V_r$ contains the first $r$ columns of $V$, $U_r$ contains the first $r$ columns of $U$, which is unique if and only if $\sigma_r > \sigma_{r+1}$. We therefore see that it is sufficient to achieve a good low rank approximation of the inverse by performing the pseudo inverse using only the largest $k$ singular values.

  From this alone it is unclear why our algorithm should work well in practice since the Hessian might not be of low rank. However, based on insights from random matrix theory and empirical evidence from large neural
  networks, we can now argue that it is sufficient for our purpose to invert only a constant number of the largest singular values.
  This means that we invert eigenvalues which lie above a certain threshold, while all eigenvalues below are replaced by a factor that is proportional to our precision limit or we set the inverse to $0$.
  We base our approximation on the insights given by Laloux \cite{laloux2000random}, Bray and Dean~\cite{bray2007statistics}, Dauphin et al.~\cite{dauphin2014identifying} and more recently Sagun et al~\cite{sagun2017empirical} as mentioned above, as well as \cite{NNRMT}.

Empirical analysis has shown that the eigenvalues-density of the Hessian matrix for a large neural network is distributed according to the Marchenko–Pastur distribution for the low eigenvalues and additionally some significantly larger outliers. The large eigenvalues in comparison are the ones which convey information about the structure of the problem and hence the direction the optimisation algorithm should move, while the continuous spectrum represents noise. Similar insights can also be found in ~\cite{sagun2016eigenvalues} and ~\cite{sagun2017empirical}, which investigate the eigenvalue distribution of the Hessian with respect to changing network architecture and inputs. The general spectrum of the Hession for large neural networks is illustrated in Figure \ref{fig:correlated} and consists of essentially three contributions:

	\begin{figure}
	\begin{center}
	  \includegraphics[width=4in]{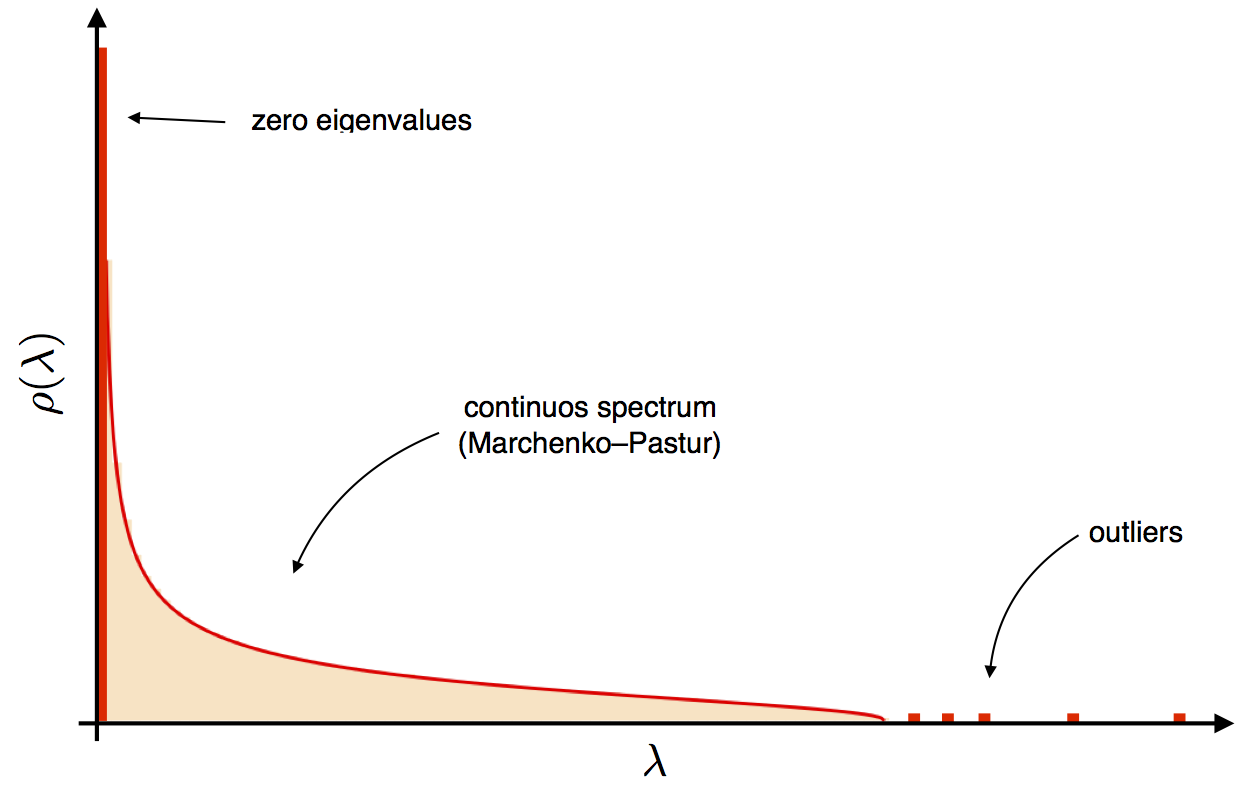}
	  \caption{Eigenvalue density of correlation matrixes.}
	  \label{fig:correlated}
	  \end{center}
	\end{figure}
	  
	\begin{itemize}
	\item Zero eigenvalues
	\item continuous spectrum (Marchenko–Pastur)
	\item Outliers
	\end{itemize}

	These observations are underpinned by recent rigorous results for spectra of correlation matrices \cite{bloemendal2016principal}. An interesting observation of \cite{sagun2017empirical} is that eigenvalues close to zero are due to the network architecture and are less and less dependent on the data as the network size (i.e.\ the depth and number of neurons) increases, while the large outliers depend mainly on the correlation structure in the data. A rough measure of the overall correlation structure of the data is given by the number of relevant principle components of the data, which we discuss more in appendix~\ref{app:pcas}. Keeping the data fixed and increasing the network size thus mainly introduces noise in the loss function but does change overall curvature structure.
	
	One can thus conclude that the eigenvalues at zero plus the continuous spectrum are not relevant for learning the curvature of the loss function as they merely represent noise. It is enough to learn the outliers, i.e. the $k$ largest eigenvalues of the Hessian. For many practical applications having $k$ of the order of $10$ to $100$ is often enough even if the dimension of the Hessian is in the hundred of thousands or millions. Furthermore by using the low rank approximation for the inverse of the Hessian we obtain a good approximation to the true Hessian (c.f. Theorem 3 in ~\cite{chung2015optimal}) which in particular conveys all important information about the curvature and hence our optimization problem.

	 This motivates the following procedure. Since only a few dominant eigenvalues exist we take only the largest eigenvalues of the Hessian when we rescale the gradient. The smallest of the large eigenvalues or some precision parameter is then taken to be the threshold, which we use to cut all other eigenvalues. Therefore we discard the rescaling of subspaces associated with noisy directions.
	 Intuitively the threshold adjusts the step size of the important directions such that they are sufficiently large, while directions which should not be taken are rescaled to a small value, proportional to the inverse of the eigenvalues. Too small eigenvalues would also lead to step sizes which take us out of the validity of the second order approximation.
	 
	 In the quantum algorithm we simply employ the above strategy by checking if the eigenvalue registers contains no information (i.e.\ all qubits are still in the $\ket{0}$ state due to the eigenvalues being smaller than the precision). Using this approximation allows our algorithm to maintain the exponential speedup when simulating the Hessian matrix and still using the important structural information for the updates. 
	 By using the absolute value, we further exploit the benefits of the saddle-free Newtons method~\cite{dauphin2014identifying} which can circumvent the problem of Newtons method being attracted to saddle-points. This is particular useful since saddle points proliferate in high-dimensional spaces~\cite{bray2007statistics,pascanu2014saddle} which act as attractors for the unmodified Newtons method.
	  
	These results hence have implications not only on our current work, but also allow for more efficient classical usage of higher order information by simply calculating the largest eigenvalues of the Hessian matrix, as was suggested previously by~\cite{dauphin2014identifying}, but in subsequent work~\cite{arjovsky2015saddle} discarded as inefficient. In particular, the above discussion combining \eqref{inverserank} with insights from random matrix theory this provides some valuable insights for classical optimization.
 
\item In the last step we ``uncompute'' the expressions and remain with the state:
  \begin{eqnarray}
    \ket{\Phi_{final}} = \sum_i (\eta_i/|\overline{\lambda}|_i) \ket{u_i}\\ \approx \ket{|H_k|^{-1}\chi}.
  \end{eqnarray}
  Following the analysis of Harrow et al.~\cite{harrow2009quantum}, 
  the error in the phase estimation procedure translates into an final error of 
  \begin{eqnarray*}
    ||\ket{\overline{|H_k|^{-1}\chi}}-\ket{|H_k|^{-1}\chi}|| = \\
    \sqrt{2 \left(1-\text{Re} \left(\braket{\overline{|H_k|^{-1}\chi}}{|H_k|^{-1}\chi}\right) \right)}\leq \epsilon,
  \end{eqnarray*}
  if we take $t=\Ord{\kappa/\epsilon}$.
  This steps requires a runtime of $\Ord{\kappa^2 \epsilon^{-3} \log{(N)}}$ since we require $\Ord{\kappa/\epsilon^3}$ copies of $\rho$.

  \item Since we need to read out the whole state including the sign of the amplitude, we further conditionally repeat the preparation of the state
  \begin{eqnarray}
    \frac{1}{\sqrt 2} \left[ \ket{0}_A \ket{+}^{\otimes p} + \ket{1}_A \ket{|H_k|^{-1} \chi} \right]
  \end{eqnarray}
  by applying the operator
  \begin{eqnarray}
    U = \frac{1}{\sqrt{2}} \left( \ket{0}\bra{0} \otimes W^{\otimes p} + \ket{1}\bra{1} \otimes \hat{U} \right),
  \end{eqnarray}
  where $W$ is the Hadamard operation \footnote{Often this is written as $H$ but we want to avoid confusion with the Hessian matrix $H$ here.} to the $\ket{+}\ket{0}^{\otimes p}$ state where $\hat{U}$ denotes the operator that prepares $\ket{|H_k|^{-1} \chi}$.
  Then apply the projection on $\ket{+}\bra{+}$, i.e.\ measurement and postselection on $\ket{+}$ in the $\pm$-basis to obtain
  \begin{eqnarray}
    \ket{+} \left[ \ket{+}^{\otimes p} + \ket{|H_k|^{-1} \chi} \right],
  \end{eqnarray}
  where $p$ is the number of qubits required for the state preparation of $\ket{H_k^{-1} \chi}$.
  Finally discard the $\ket{+}$ register and apply a projection into the computational basis $\ket{j} \bra{j}$ gives
  \begin{eqnarray}
    P_{\ket{j}\bra{j}} \left[ \frac{\ket{+}^{\otimes p} + \sum_i \alpha_i \ket{i}}{\sqrt{2}} \right]
  \end{eqnarray}
  where $\sum_i \alpha_i \ket{i} \equiv \ket{|H_k|^{-1}\chi}$ and we end up with the probability to measure $j$ to be
  \begin{eqnarray}
   \left(\frac{1}{\sqrt{2^p}} + \alpha_j \right)^2.
  \end{eqnarray}
  To obtain a good estimate of this we require $\Ord{N \log{N}}$ measurements and $\Ord{2n}$ qubits.
  This leads then to a total of $\Ord{\kappa^2 \epsilon^{-3} N \log^2 N}$ for this step.
\end{enumerate}

\begin{algorithm}[t]
  \caption{Algorithm for obtaining the step-update of the truncated-Newton method}
  \label{algo2}
  \begin{enumerate}
    \item Prepare $\ket{\bsy \nabla f} = \sum_i \eta_i \ket{v_i}$ with $v_i$ being the singular vectors of $H$ and hence the eigenvectors, since $H$ is symmetric.
    \item Prepare multiple copies of $\rho_{HH^{\dagger}}$ and perform the matrix-exponentiation method to \textit{apply} 
    the Hessian to the state $\ket{\bsy \nabla f}$, followed by the phase estimation routine.
    \item Apply a controlled rotation $R_Y\sin^{-1}(1/\bar{\sigma}_i))$, where $\bar{\sigma}_i=|\bar{\lambda}_i|$, with 
    $\bar{\lambda}_i$ being the approximate eigenvalues of the Hessian matrix.
    \item Perform postselection on $\ket{1}$ and uncompute the ancilla registers to obtain the state
    \begin{eqnarray}
    \ket{\Phi} &=& \sum_i (\eta_i/|\overline{\lambda}_i|) \ket{v_i}  \nonumber \\ &\approx& \ket{|H|^{-1} \bsy \nabla f}.
    \end{eqnarray}
    \item If the classical state needs to be recovered, repeat the process with the state 
    $\left[ \ket{+}^{\otimes n} + \ket{|H|^{-1} \bsy \nabla f} \right]$ and obtain good statistics, which will require $\Ord{N\log{N}}$ time.
  \end{enumerate}
\end{algorithm}

We summarize the complete algorithm in the box Algorithm \ref{algo2}. In order to perform the last step we need to be able to conditionally prepare the state. This can be done by deferring the measurement till the end of the simulation. Neglecting the details of the actual algorithm this can be trivially done by preparing the state
\begin{eqnarray}
  \frac{1}{\sqrt{2}} \ket{+}^{\otimes n} \otimes \ket{1}_C + \frac{1}{\sqrt{2}} \sum_i C_i \ket{\lambda_i} \left( \sqrt{\left( 1- \frac{1}{\lambda_i^2}\right)} \ket{0}_C + \frac{1}{\lambda_i} \ket{1}_C \right)  \nonumber \\ =
   \frac{1}{\sqrt{2}} \left( \ket{+}^{\otimes n} + \sum_i C_i \frac{1}{\lambda_i} \ket{\lambda_i} \right) \otimes \ket{1}_C  + \nonumber 
   \frac{1}{\sqrt{2}} \sum_i C_i \sqrt{\left( 1- \frac{1}{\lambda_i^2}\right)} \ket{\lambda_i} \otimes \ket{0}_C ,
\end{eqnarray}
and then performing the postselection on the $\ket{1}_C$ state.

The classical truncated Newton's method has empirically been demonstrated by Martens~\cite{martens2010deep}, and outperformed standard methods like gradient descent with respect to the number of iterations and the obtained error. In particular, truncated Newton's method is still able to find the minimum in optimization problems which are hard for standard gradient descent methods as for example the Rosenbrock function. Further it circumvents many problems of curvature blind methods like gradient descent, in particular for cases of pathological curvature~\cite{martens2010deep} while still maintaining a quadratic runtime for the exact and almost linear runtime for the approximate case.

Assuming dense Hessian matrices, we find that our algorithm has a quadratically better dependency on the parameter dimensions while obtaining a worse dependency on the condition number and precision. We yet expect that the condition number dependency can be improved based on known results~\cite{clader2013preconditioned}. Furthermore, as we discussed above, our method automatically bounds the condition number of the matrix since we truncate the lowest eigenvalues. This procedure hence automatically introduces an upper bound of the condition number that is given by $1/[\text{threshold}]$.

Our quantum-classical hybrid method compares hence to known methods like quasi-Newton or Hessian-free methods, but is expected to have a better performance and convergence time, since it uses the exact (largest) eigenvalues and runs still in linear time with respect to the dimension of the problem.

\section{Discussion}

We presented a quantum-classical hybrid algorithm which can be used to efficiently train deep neural networks and auto encoders. Our algorithm improves upon other classical algorithms as the problem allows for the efficient integration of quantum subroutines. More precisely, we achieve a polynomial (quadratic) speed-up in the problem dimensionality, i.e.\ the network parameters, and expect further a faster convergence rate of the algorithm. Note that many quantum algorithms either provide a quadratic improvement, e.g. Grover's search or a exponential improvement, e.g. quantum Fourier transform. While some modern quantum machine learning algorithms such as Harrow et al.~\cite{harrow2009quantum} matrix inversion state an exponential speedup, such a speedup cannot be achieved when a classical readout of the result is required, but rather can only be achieved if this algorithm is used as a subroutine where the final quantum state is passed on to the next routine. A similar situation applies to our algorithm where our speed up is limited by the fact that a classical readout of our final result already takes $\mathcal{O}(N)$. If one were able to extend the algorithm to a purely quantum algorithm in lights of Rebentrost et al.~\cite{rebentrost2016quantum} and Kerenidis and Prakash~\cite{kerenidis2017quantum} one could in principle achieve an exponential speedup.

In the development of our algorithm we present arguments that explain the success of the truncated (saddle-free) Newton's method based on empirical and analytical results. In particular, combining Theorem 3 of Chung et al.~\cite{chung2015optimal} with empirical analyses from high-dimensional optimisation and deep neural networks \cite{martens2010deep,pascanu2014saddle,dauphin2014identifying,sagun2016singularity,sagun2017empirical} as well as recent theoretical insights from random matrix theory \cite{bloemendal2016principal}, provides new insights into the success of the truncated (saddle-free) Newton's method. Relating \eqref{inverserank} with Figure \ref{fig:correlated} explains why it suffices to only consider the largest eigenvalues of the Hessian in the (saddle-free) Newton's method. In fact, introducing more and more eigenvalues only introduces more noise in the curvature and can potentially even harm the optimization procedure. Our approach may inspire a new way to design quantum-classical hybrid algorithms that make use of specific traits inherit in quantum subroutines in order to improve classical state-of-the-art algorithms. 

{\it Acknowledgement --} The authors thank Simon Benjamin and Philipp Henning for their kind support and helpful discussions. SZ would also like to thank Max Atkin, Luka Skori\'c and Steve Roberts for fruitful discussions especially during the work on \cite{NNRMT}.

\bibliographystyle{apsrev4-1}
\bibliography{semi_quantum}

\appendix

\section{Data principal components and the Hessian eigenvalue distribution}
\label{app:pcas}

\begin{figure}[!htp]
  \centering
  \begin{minipage}[b]{0.4\textwidth}
    \includegraphics[width=\textwidth]{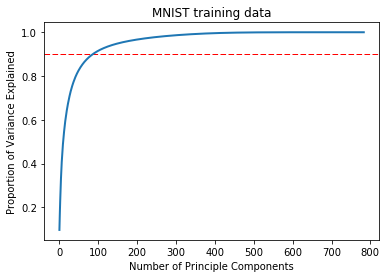}
	 \caption{PCA of MNIST}  
     \label{fig:pcas}
     \end{minipage}
  \hfill
  \begin{minipage}[b]{0.4\textwidth}
    \includegraphics[width=\textwidth]{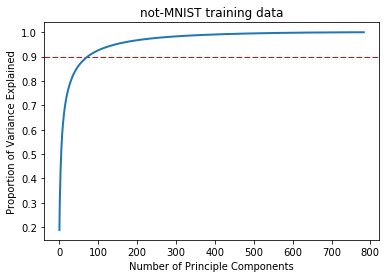}
  	\caption{PCA of not-MNIST}
    \label{fig:pcas_2}
  \end{minipage}
\end{figure}

In this section we briefly summarize insights from various sources in order to support our above outlined procedure.
Bloemendal et al.~\cite{bloemendal2016principal} analyzed the principal components of the sample covariance matrices and found that the spectrum is reminiscent of the description we gave in section~\ref{sec:quantum_truncated} for the Hessian matrix.
Sagun et al.~\cite{sagun2017empirical} then showed that the number of eigenvalues of the Hessian which are not in the bulk nor singular depend less on the network size but are more dependent of the structure of the data. This leads to the conclusion that the number of significant eigenvalues of the Hessian in the training of deep neural networks is mainly depended on the correlation structure of the input data rather then the network size. A good proxy for the amount of correlation structure in the input data is the number of principle components above a certain threshold, e.g.$90\%$, which is of the same order as the significant eigenvalues of the Hessian matrix \cite{sagun2017empirical}. We performed a similar analysis for the MNIST and non-MNIST data set and found that our results support this claim. The number of principal components which cumulatively explain $90\%$ of the data variance is close to the number of outliers in the eigenvalue spectrum. The proportion of variance explained is shown in figure~\ref{fig:pcas} for MNIST and in~\ref{fig:pcas_2} for not-MNIST data. We observe that in fact only a fraction of the principal components of the data set is relevant to explain almost the entire variance.

Based on various works~\cite{sagun2016singularity,sagun2016eigenvalues,sagun2017empirical} which suggest that the number of outliers in the spectrum of the Hessian matrix is only depending on the data set and not on the network size and on the insight that the data variance is explained by a few principal components we hypothesize that there is a direct connection between the number of significant eigenvalues of the Hessian and the number of principal components of the data. As the latter is independent of the network specifications we believe that this is further strong evidence for the validity of our algorithms.

\begin{figure}
\begin{center}
  \includegraphics[width=3in]{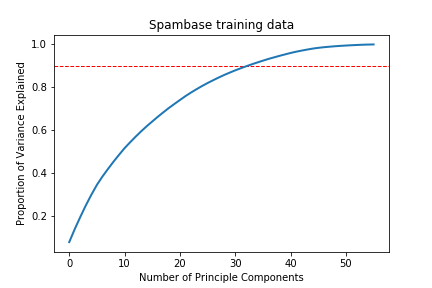}
  \caption{PCA for Spambase}
  \label{fig:pca_spambase}
  \end{center}
\end{figure}

The corresponding proportion of variance explained for the Spambase set is given in Figure~\ref{fig:pca_spambase}. We see therein, that the behavior is slightly different to the MNIST and Not MNIST data set, but can also be explained with a small number of singular values.

\section{Calculations for density matrix exponentiation}
\label{app:density_matrix}

We briefly demonstrate the quantum principal component analysis calculus. We thereby demonstrate the equality between the repeated swap operator
application to the tensorproduct state $(\rho \otimes \sigma)$ and the time evolution. We use the following definitions:
\begin{itemize}
  \item $\rho = \sum_{i,j} \alpha_{ij} \ket{i}\bra{j}$
  \item $\sigma = \sum_{k,l} \beta_{kl} \ket{k}\bra{l}$
  \item $S = \sum_{m,n} \ket{m}\bra{n} \otimes \ket{n}\bra{m}$
\end{itemize}

\begin{enumerate}
  \item We start by demonstrating the one direction first:
   \begin{eqnarray*}
    \trc{1}{e^{-iS\Delta t} \left(\rho \otimes \sigma \right) e^{iS\Delta t}} =\\
    \trc{1}{ (Id - iS \Delta t + \Ord{\Delta t^2} ) \left(\rho \otimes \sigma \right) ( Id + iS \Delta t + \Ord{\Delta t^2} )} = \\
    \trc{1}{ \left(\rho \otimes \sigma \right)  - iS \left(\rho \otimes \sigma \right) \Delta t  + iS \left(\rho \otimes \sigma \right)  \Delta t + \Ord{\Delta t^2}} =\\
    \sigma i - \Delta t [\sigma, \rho] +\Ord{\Delta t^2};
  \end{eqnarray*}
  \item We further show explicitly the trace operation on the mixed terms:
  \begin{eqnarray*}
    \trc{1}{ (S \left(\rho \otimes \sigma \right) } = 
    \trc{1}{\sum_{m,n} \sum_{i,j} \sum_{k,l} \alpha_{ij} \beta_{kl} \bra{m}\ket{n}  \ket{i}\bra{j} \otimes \bra{n}\ket{m} \ket{k}\bra{l} } =\\
    \trc{1}{\sum_{i,j} \sum_{k,l} \alpha_{ij} \beta_{kl} \ket{k}\bra{j} \otimes \ket{i}\bra{l} } = 
    \sum_r \sum_{i,j} \sum_{k,l} \alpha_{ij} \beta_{kl} \bra{r}\ket{k}\bra{j}\ket{r} \otimes \ket{i}\bra{l} =\\
    \sum_r \sum_{i,j} \sum_{k,l} \alpha_{ij} \beta_{kl} \delta_{rk} \delta_{jr} \ket{i}\bra{l} = 
    \sum_{r,i,l} \alpha_{ir} \beta_{rl} \ket{i}\bra{l}  = \rho \cdot \sigma,
  \end{eqnarray*}
  since $\rho \cdot \sigma = \sum_{r,i,l} \alpha_{ir} \beta_{rl} \ket{i}\bra{l}$ and 
  $\sigma \cdot \rho = \sum_{r,i,l} \alpha_{ri} \beta_{lr} \ket{l}\bra{i} $;
  \item Compare this to:
  \begin{eqnarray*}
    e^{-i \rho\Delta t} \sigma e^{i \rho \Delta t} = 
    (Id - i \rho \Delta t + \Ord{\Delta t^2} ) \sigma  ( Id + i \rho \Delta t + \Ord{\Delta t^2} ) = 
    \sigma - i \Delta t (\rho \sigma - \sigma \rho) + \Ord{\Delta t^2}.
  \end{eqnarray*}
\end{enumerate}

In our algorithm we need to be able to prepare the density matrix encoding the $HH^T$ or efficiently. We start by demonstrating a method to prepare the following density matrix, using the following oracles:
\begin{eqnarray}
 U_f \ket{i,j} \ket{0}= \ket{i,j}\ket{H_{ij}} \\
 U_g \ket{i} \ket{0}= \ket{i}\ket{(\nabla f)_i} .
\end{eqnarray}
We can apply the oracles to the density matrix $\rho_0$, so that we find
\begin{align*}
\rho _1= U\rho _0\otimes |0\rangle \langle 0|U^{\dagger } = \sum_{j}U|\tilde{\psi
  }_j\rangle \langle \tilde{\psi} _j|\otimes |0\rangle \langle 0|
  U^{\dagger } = \sum_{j,i,i'}\alpha _{i,j}\alpha ^*_{i'j}|i,j\rangle \langle i',j|
|H_{ij}\rangle \langle H_{i'j}|,
\end{align*}
where we used
$ U|\tilde{\psi }_j\rangle |0\rangle =\sum_{i}\alpha _{i,j}|i,j\rangle
|H_{ij}\rangle $ and $ \langle \tilde{\psi }_j| \langle 0|U^{\dagger }
=\sum_{i'}\alpha^* _{i'j}\langle i',j|
\langle H_{i'j}| $.\\

 The initial state is given by $ (1/\sqrt{N})\sum_{i,j}\alpha _{i,j}\ket{i,j}$ 
where $\alpha$ can be taken to be $1/N$ if we can e.g.\ use the state $H^{\otimes (n+n)} \ket{0}^{\otimes (n+n)}$ as input.
The density matrix of this state is then given by
\begin{eqnarray}
  \rho =\frac{1}{N}\sum_{i',j',i,j}\alpha _{i,j}\alpha ^*_{i',j'} \ket{i,j} \bra{i'j'}.
\end{eqnarray}
We then dephase this state using the Hamiltonian
$H_{{\text{noise}}}= \pi f(t)\sum_{j} |j\rangle \langle j|$, where $f(t)$ is a classical random variable (which changes over time)~\cite{matsuzaki2011magnetic}, and $\pi$ is a coupling constant. We then obtain
\begin{eqnarray}
 \rho _0=\frac{1}{N}\sum_{i',i,j}\alpha _{i,j}\alpha ^*_{i',j} \ket{i}.
  \bra{i'}\otimes \ket{j} \bra{j}.
\end{eqnarray}
We next define the state $\ket{\tilde{\psi }_j} =\sum_{i}\alpha _{i,j}\ket{i,j}$ and with this the density matrix
\begin{eqnarray}
 \ket{\tilde{\psi }_j} \bra{\tilde{\psi} _j}=\sum_{i,i'}
  \alpha _{i,j}\alpha _{i',j}^* \ket{i,j} \bra{i',j}.
\end{eqnarray}
We can use this to rewrite $\rho _0$ as
\begin{eqnarray*}
 \rho _0=\frac{1}{N}\sum_{i',i,j}\alpha _{i,j}\alpha ^*_{i',j} \ket{i}
  \bra{i'}\otimes \ket{j} \bra{j} = \sum_{j}|\tilde{\psi
  }_j\rangle \langle \tilde{\psi} _j| =
  \sum_{j}p_j|\psi _j\rangle \langle \psi _j|
\end{eqnarray*}
where $p_j =1/{\text{ Tr} }[|\tilde{\psi }_j\rangle \langle \tilde{\psi}
_j|]$ and $|\psi _j\rangle \langle \psi _j|=|\tilde{\psi }_j\rangle \langle \tilde{\psi}
_j|/{\text{ Tr} }[|\tilde{\psi }_j\rangle \langle \tilde{\psi}
_j|] $.

 We next add an additional ancilla qubit and apply a controlled $R_Y$-rotation depending on the value of the $H$-register: 
\begin{eqnarray*}
 \rho _2=U_{CR} (\rho _1\otimes |0\rangle \langle 0|)U^{\dagger }_{CR} =\sum_{j}U_{CR}U(|\tilde{\psi
  }_j\rangle \langle \tilde{\psi} _j|\otimes |00\rangle \langle 00|)
  U^{\dagger }U_{CR}^{\dagger }
\end{eqnarray*}
where we have 
$$  U_{CR}U|\tilde{\psi
  }_j\rangle |00\rangle  =\sum_{i}\alpha _{i,j}|i,j\rangle
  |H_{ij}\rangle \left(\sqrt{1-\left(\frac{H_{ij}}{c}\right)^2}|0\rangle
  +\frac{H_{ij}}{c}|1\rangle \right)
$$
and
$$    \langle \tilde{\psi
  }_j| \langle 00|U^{\dagger}_{CR}U^{\dagger } =\sum_{i'}\alpha^*
  _{i',j}\langle i',j| \langle H_{i'j}|
  \left(\sqrt{1-\left(\frac{H_{i'j}}{c}\right)^2}\langle 0| +\frac{H_{i'j}}{c}\langle 1| \right),
$$ and we defined the controlled rotation to act as:
\begin{eqnarray}
 U_{CR}=|H_{ij}\rangle |0\rangle =|H_{ij}\rangle
  \left(\sqrt{1-\left(\frac{H_{ij}}{c}\right)^2}|0\rangle +\frac{H_{ij}}{c}|1\rangle \right).
\end{eqnarray}

In the last step we perform a measurement on the ancilla qubit and postselect on $\ket{1}$ and finally uncompute the ancilla registers.
That means we project the qubit into $\ket{1} $:
  \begin{eqnarray}
   \tilde{\rho }_3=|1\rangle \langle 1|\rho _2 |1\rangle \langle 1|  =\sum_{j}|1\rangle \langle 1|U_{CR}U(|\tilde{\psi
  }_j\rangle \langle \tilde{\psi} _j|\otimes |00\rangle \langle 00|)
  U^{\dagger }U_{CR}^{\dagger }|1\rangle \langle 1| =\sum_{j}|\tilde{\phi
  }_j\rangle \langle \tilde{\phi}_j| ,
  \end{eqnarray}
  where we ignored the renormalization and we have $$|\tilde{\phi }_j\rangle
  =|1\rangle \langle 1|U_{CR}U|\tilde{\psi
  }_j\rangle |00\rangle =\sum_{i}\alpha _{i,j}\frac{H_{ij}}{c} |i,j\rangle
  |H_{ij}\rangle |1\rangle . $$
  This is followed by the sol-called ``uncompuation'' step 
  \begin{eqnarray}
   \rho _4=U^{\dagger }\rho _3U=\sum_{j}U^{\dagger }|\tilde{\phi
  }_j\rangle \langle \tilde{\phi}_j|U
  \end{eqnarray}
  where
  $$U^{\dagger }|\tilde{\phi }_j\rangle
  =\sum_{i}\alpha _{i,j}\frac{H_{ij}}{c}U^{\dagger } |i,j\rangle
  |H_{ij}\rangle |1\rangle =\sum_{i}\alpha _{i,j}\frac{H_{ij}}{c} |i,j\rangle
  |0\rangle |1\rangle  $$ and $$\langle \tilde{\phi }_j|U=\sum_{i'}\alpha^*
  _{i',j}\frac{H_{i'j}}{c}\langle i'j|\langle 0|\langle 1|.$$

  Therefore we obtain
  \begin{eqnarray}
  \label{dm_prep}
   \rho _4 &=&\rho _5 \otimes |01\rangle \langle 01|\nonumber \\
   \rho _5&=&\sum_{i,i'}\alpha _{i,j}\alpha
    ^*_{i',j}\frac{H_{ij}H_{i'j}}{c^2}|ij\rangle \langle i'j|
  \end{eqnarray}
  and since $H_{ij}=H_{ji}$, we have
    \begin{eqnarray}
    \rho _5&=&\sum_{i,i'}\alpha _{i,j}\alpha
    ^*_{i',j}\frac{H_{ij}H_{ji'}}{c^2}|ij\rangle \langle i'j|.
  \end{eqnarray}
  By finally tracing out the system for $j$, we obtain
    \begin{eqnarray}
    \label{final_density}
    \rho '_5&=&\sum_{i,i',j}\alpha _{i,j}\alpha
    ^*_{i',j}\frac{H_{ij}H_{ji'}}{c^2}|i\rangle \langle i'|.
  \end{eqnarray}

\section{Sample based SVD}
\label{app:sample_svd}

\begin{theorem}[Linear time approximate SVD~\cite{drineas2006fast}]
\label{thm:approx_svd}
Suppose $A \in \mathbb R^{m \times n}$ and let $H_k$ be constructed by the $\Ord{c^3 + c^2m}$-time algorithm LINEAR\_TIME\_SVD in~\cite{drineas2006fast} by sampling $c$ columns of $A$ with probabilities $\{p_i \}_{i=1}^n$ such that $p_i \geq \beta |A^i|^2/||A||_F^2$ for some positive $\beta \leq 1$, and let $\eta =  1+\sqrt{(8/\beta) \log(1/\delta)}$. Let $\epsilon >0$. If $c \geq 4k/\beta \epsilon^2$, then 
\begin{eqnarray}
\mathbb E [||A - H_k H_k^T A||_F^2] \leq || A - A_k ||_F^2 + \epsilon ||A||_F^2,
\end{eqnarray}
where $A_k$ is the best rank-$k$ approximation to $A$.
If $c \geq 4 k \eta^2/ \beta \epsilon^2$, then with probability at least $1-\delta$, 
\begin{eqnarray}
||A - H_k H_k^T A||_F^2 \leq || A - A_k ||_F^2 + \epsilon ||A||_F^2.
\end{eqnarray}
In addition if $c \geq 4/ \beta \epsilon^2$, then 
\begin{eqnarray}
\mathbb E [||A - H_k H_k^T A||_2^2] \leq || A - A_k ||_2^2 + \epsilon ||A||_F^2,
\end{eqnarray}
and if $c \geq 4 \eta^2/ \beta \epsilon^2$, then with probability at least $1-\delta$,
\begin{eqnarray}
||A - H_k H_k^T A||_2^2 \leq || A - A_2 ||_F^2 + \epsilon ||A||_F^2.
\end{eqnarray}
\end{theorem}

\end{document}